# Motion and audio analysis in mobile devices for remote monitoring of physical activities and user authentication

Hamed Ketabdar*, Jalaluddin Qureshi[#], Pan Hui*.

*Quality and Usability Lab, Deutsche Telekom Laboratories, Berlin, Germany.
[#]School of Computer Engineering, Nanyang Technological University, Singapore.

hamed.ketabdar@telekom.de, jala0001@e.ntu.edu.sg, pan.hui@telekom.de

# Motion and audio analysis in mobile devices for remote monitoring of physical activities and user authentication

In this paper we propose the use of accelerometer, embedded by default in smartphone, as a cost-effective, reliable and efficient way to provide remote physical activity monitoring for the elderly and people requiring healthcare service. Mobile phones are regularly carried by users during their day-to-day work routine. A user's physical movement information during the user's daily routine can be captured by the mobile phone accelerometer, processed and sent to a remote server for monitoring. The acceleration pattern can deliver information related to pattern of physical activities the user is engaged in. This information can also be used by a remote monitoring agent to monitor user's health related information. We also show how this technique can be extended to provide implicit real-time security by analysing unexpected movements captured by the phone accelerometer, and automatically locking the phone in such situation to prevent unauthorised access. This technique is also shown to provide implicit continuous user authentication, by capturing regular user movements such as walking, and requesting for re-authentication whenever it detects a non-regular movement.

## 1. Introduction

Smartphones are becoming more and more popular among consumers, supporting a wide array of smart applications and utilities which can run on them. An interesting number of such applications are now based on sensors (Lane *et al.* 2010) such as accelerometer, gyroscope, dual cameras, GPS, digital compass and dual microphones which are implicitly embedded in an increasing number of smartphones. Such growth in the number of sensor based smart applications has been supplemented due to development of smaller Integrated Circuits (IC) and an increasing computational and memory capacity technology available for smartphone. This means that more user functionalities can be added in smartphone, while keeping the size of the device smaller. Popular smartphone Operating Systems (OS) such as Android, Symbian and iOS allow third-party application development using Software Development Kit (SDK), making it much easier for vendors to custom design and distribute their own software applications (freeware/ commercial) through the online app store of the smartphone vendor.

      Taking advantage of these hardware development and open third party application development, we propose a novel smartphone application as a mean to remotely monitor the physical activity of the user, using movement information detected from the smartphone's accelerometer. Previous studies (Karantonis *et al.* 2006) on monitoring the human kinesiology were based on the use of specially designed external sensors, which had to be explicitly carried by the users. Such sensors were also limited in its scope of target audience as they were primarily designed for those users requiring healthcare administration. These external wearable sensors were expensive, "inconvenient" to be carried during day-to-day work routine, and were targeted for a specific category of users, which narrowed down their market scope, and therefore kept the cost of these sensors high.

Accelerometer sensors have already been successfully used in several other applications, especially for sensing device orientation. Rekimoto (2001) discussed the potential of this technique for tasks such as navigating menus and scrolling. Hinckley *et al.* (2000) demonstrates how accelerometers could be useful for an automatic screen orientation device and scrolling application. Oakley and O'Modhrain (2005) describe a tilt based system with tactile augmentation for menu navigation.

As smartphones become an increasingly popular consumer device, regularly carried by the users, it offers a more pragmatic solution to monitoring the human physical activities for a wide range of purposes. The convenience, with which smartphone software application can be written to suit the requirements of individual users, also makes it more appealing to wider user audiences, and not just for those requiring healthcare administration. In this paper we propose a physical activity measurement application for smartphone user. Such physical activity monitoring system therefore may not be just limited only to users requiring healthcare administration, but even healthy individuals can also benefit from such application.

We then also show how this technique can be extended to provide security to mobile device, by using the user movement patterns and audio information. We propose a security paradigm which allows for online, implicit and continuous protection of data without the burden of involving the active attention of the user. We show that analysis of audio and physical movement data captured by a mobile phone can be used to indicate unexpected events when it occurs. Such unexpected event can possibly occur due to the phone being lost or stolen, when the phone has been left unattended for a long duration or when the phone hits a hard surface from a short height. In addition, we show that analysis of audio and physical movement data during regular physical activities (e.g., walking, working in the office) by the user can allow for automatic authentication of the user. The proposed method is an implicit authentication technique, i.e., it does not involve active attention of the user, and it is performed continuously as the user is regularly using or carrying the device. Physical movement data can be captured by the accelerometer sensor embedded in modern mobile devices. Audio data is captured by embedded microphone.

Social implication of such applications applies to a wide range of users. Activity monitoring application can be used remotely to provide healthcare administration to a user, without requiring the physical presence of healthcare staff with the patient. With an ageing population, integration of such technologies can reduce the manpower and financial strain on the healthcare sector. Whereas the use of movement and audio information to detect unexpected events, lock the phone and request re-authentication discourage mobile phone theft. We will further discuss the social implications of our proposed applications in the subsequent section.

In the work we present our approach for analysing activity pattern of a user using acceleration data captured by the mobile phone, and its applications in monitoring and assisted life. In this work, we are mostly interested in monitoring the amount of user activity, as well as classifying user activities into certain groups. We also show how this work can be extended to provide implicit security features during unexpected events. We further show that in addition to movement information, audio information can also be used to provide a higher level of security to the mobile phone by automatically locking when the phone senses unfamiliar movement and audio information.

The rest of the paper is organised as follow. In Section 2 we discuss various applications and scenarios where movement analysis information can be useful. In Section 3, we present signal processing techniques used to interpret movement data from the accelerometer for the purpose of estimating the user activity level and activity classification. In Section 4, we then give results of our experimental and user studies conducted to evaluate the performance of our activity monitoring application system. In Section 5 we present ActivityMonitor, which is a system we developed to remotely measure the physical activity of a user, and its social implication. We then provide conclusion of our work in Section 6.

**2. Practical uses of accelerometer based applications**

Information from smartphone accelerometer can be creatively used to design new smartphone application based on human movement. These applications can be developed by third-party, and distributed through vendor online app store. This also does not impose any hardware changes, as accelerometers are implicitly embedded in the latest generation of smartphones. An accelerometer detects the rate of change of velocity. Common human movement such as walking results in changes of velocity, which can be detected by accelerometer.

Different people have different walking patterns (such as average speed, acceleration and deceleration). These different patterns extend to other daily movement behaviours such as walking on the stairs, waiting for the bus and watching television while seated on a sofa and so on. Each user will have his unique set of movement patterns. Since users carry their mobile phones with them all the time, information from these movement pattern can be used for monitoring the activity of the user for remote healthcare administration. Further, since these movement patterns are uniquely associated with different users, information from these movements can also be used to provide security, by automatically locking the phone, when the phone senses unfamiliar movement or situation. In this section we demonstrate how movement pattern information can be used for a variety of useful application, particularly for activity monitoring and unexpected event detection.

*2.1 Activity monitoring*

While remote activity monitoring can particularly benefit patients requiring remote monitoring, such activity monitoring application can also be used for a host of other purposes. Athletes interested to keep track of their running pattern (e.g. speed variation over the course of marathon running) may use a physical activity application which will record their acceleration patterns. Security guards working in small group (or worse, working alone!) can benefit from an application which will send an alarm message for assistance if the application does not detect any movement over a short duration, say 5 minutes for example. Game developers can use an accelerometer and gyroscope based physical activity detection application for designing games which will require the physical movement of the user to play the games. Healthy people can benefit from application which periodically sends the user report detailing his activity level and energy consumption, which the user can use to watch his daily calories requirement. The possibilities to creatively use such physical activity application are therefore only limited by the designer's creativity.

Most importantly this remote physical activity monitoring system can benefit the healthcare sector. Aging population in Western Europe and industrialised countries such as Japan, South Korea and Canada is increasing. This increase in the ratio of dependents per working person in many economies clearly puts the healthcare sector under both workforce and financial strain. Remote healthcare administration has gained prominence as a viable solution to remotely monitor the patient, and therefore has the prospect to offset both the clinical manpower and financial strain imposed due to the aging population.

Various projects have already been initiated to benefit the healthcare sector by integrating Information and Communication Technology (ICT) solution to provide smarter healthcare administration to patients. The SAPHE project, a research collaboration between academia and industry has been working on to provide Telecare and Telehealth applications using wireless sensors. The E-care project aims to provide general social mobile welfare system provision and alarm message to nurses when these patient need special attention. This way it reduces continuous physical monitoring of the patient by home based health care providers.

Regarding the issue of heath care and activity analysis, there are a few mobile phone based commercial applications especially for iPhone platform (Runkeeper, MapmyRIDE). These applications use data provided by GPS position signal to analyse amount of activity during sports. However, these applications are dependent on availability of GPS signal for activity level estimation. In many places (such as home), GPS signal is not available due to coverage problems. In addition, limited and mostly expensive mobile phones are equipped with GPS receivers, while acceleration sensor integration in mobile phones is more widespread and much less expensive. After all, user position information is not strongly correlated with amount or pattern of user activity. In contrast, acceleration information is strongly correlated with amount and pattern of force exerted by user, and thus more directly related to amount and pattern of user activity.

Custom-designed terminal such as the Motorolla MC75A0-HC Healthcare Terminal is primarily designed for use by the nursing staff to administer healthcare for the patient, it doesn't address the need to monitor the patient remotely. Our model runs on top of the accelerometer already embedded in smartphone by default. Therefore our design does not suffer from any changes in smartphone hardware, and the need to carry any special purpose device such as a sensor by the user. We propose to use data provided by mobile phone's accelerometer for examining activity pattern of its user. The mobile phone equipped with these sensors is normally carried by the user in his pocket (Ichikawa *et al.* 2005). The results of examination can be presented to the user or a monitoring agent in different ways as indications of different health related factors. In addition to presenting these data to the user, the mobile phone can then optionally analyse these data, or send it to a server for further analysis. The mobile phone or server can analyse physical activity pattern of the user and compare it against normally accepted pattern for the same user, or normally accepted activity pattern for the users of the same age.

## 2.2 Unexpected event detection

Unexpected events experienced by a mobile device can be a sign of security threats, and upon detecting such scenario, the phone can be automatically locked, requiring user re-authentication. This way such unexpected event detection application can be used to prevent anyone else from accessing the user's confidential information stored in the phone and increase the security of the phone from being stolen and protecting user confidential information. Consider a simple case where the mobile phone accelerometer has not detected motion for a relatively long period of time, it may indicate that the phone is lost or forgotten somewhere. This may result in a security risk for data or services accessed by the phone. This situation can be easily identified by analyzing motion data obtained from the embedded acceleration sensors. In this case, the rate of change of acceleration data can be quite low over a long period of time. Upon detection of such a situation, the device can be automatically locked and request user re-authentication.

Another risky situation can be when the user is engaged in high level of physical activity (e.g., running). In such a situation, attention of user to his mobile phone may be reduced, resulting in a relatively higher risk of losing the device. Detecting such a situation based on embedded accelerometer sensors allows setting security options on the mobile device to a higher level. According to our experiments (assuming that the mobile device is carried somewhere on the user's body), the average and variance of acceleration show significantly high values in these cases, as compared to regular situations.

A common risky situation which can lead to having the device lost or stolen is when the mobile device falls out of user's pocket or bag, and is left unattended for a while (Fig.1). The user may not notice what has happened to the device due to distractions with other activities. We have studied detecting such a scenario based on analysing audio and acceleration information. We model this unexpected event based on three sub-events: free-fall, shock (impact with floor), and no activity (movement) after the shock. When the phone falls, it initially experiences a free fall situation. Upon hitting the floor, it experiences a shock. However, this is not enough for identifying the situation as "risky", because the user may immediately pick up the phone. Therefore, it is also necessary to check for a period of "no activity" after the shock.

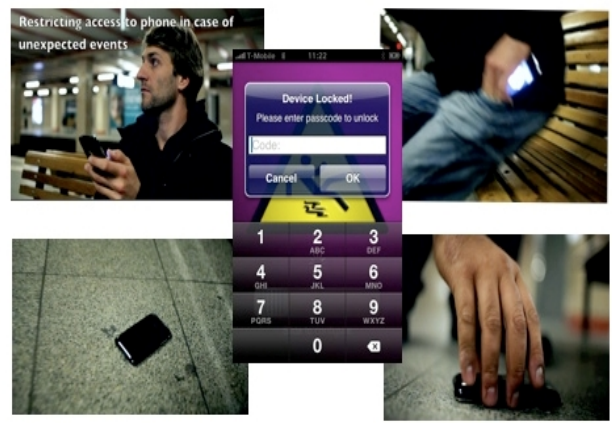

Figure 1. A mobile phone under risk of being lost or stolen

## *2.3 Implicit user identification based on audio and movement pattern analysis*

In this section we consider a more resilient security scheme. In addition to using the user movement pattern, information from the phone microphone can also be used to provide better security for the phone. Audio and movement analysis can be used to enhance the security functionalities in mobile devices for user identification based on user's regular physical activities such as walking. When the mobile device is carried by the user (e.g., in his pant pocket), it can capture samples of audio and motion information, and check for a biometric patterns in them. In this way, the identity of the user can be verified in a continuous and implicit manner. The authentication is implicit, so the user does not need to actively participate in authentication process. The user only performs his regular activities and the authentication method looks for a biometric sign in his pattern of physical activities. Once the device automatically detect that it is not being carried or operated by the same user anymore, it can switch to a higher level of required authentication. As a side advantage, this technique reduces the required number of re-authentications. If the device implicitly detects that it has been continuously used by the same person since the last authentication, it may not ask for a new authentication process for the same service. This reduces the number of repetitive re-authentication. In addition, an implicit authentication score estimated based on audio and movement analysis can be used to set up different security threat levels for the mobile device, allowing implementation of a graded security scheme.

**3. Processing the movement patterns**

In this section we provide technical details of how the movement pattern data is analysed and preprocessed to remove "acceleration noise." Once preprocessed, useful features are extracted from the signal, and this information is then used for estimating the activity level of the user and for user activity classification.

*3.1 Analysis of acceleration patterns*

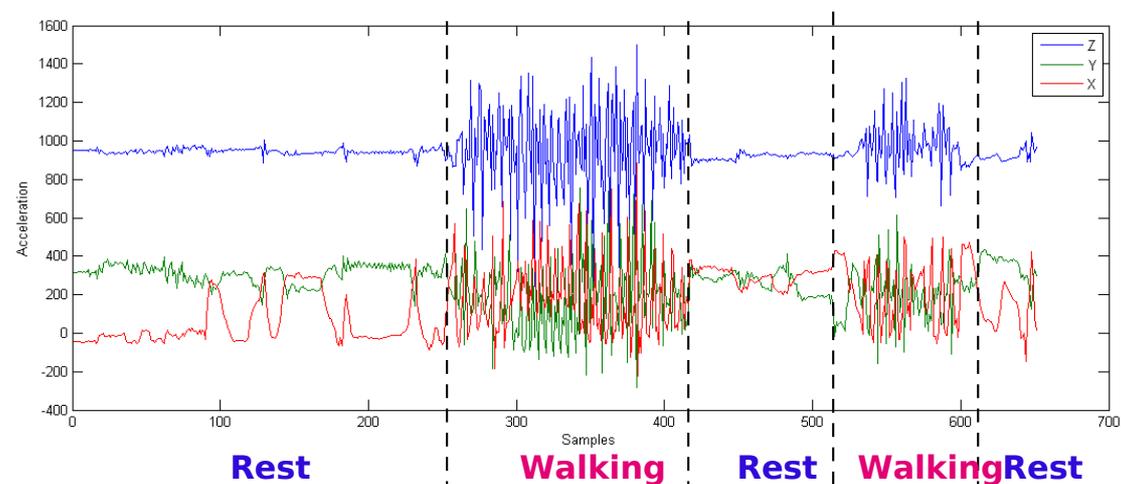

Figure 2. An example of acceleration signals captured by acceleration sensors during walking and resting activities.

Acceleration sensors integrated in a mobile phone provide linear acceleration information along the x, y, and z axis. These sensors were originally used for automatic screen rotation and navigation (Rekimoto 2001, Rekimoto 1996, Hinckley *et al.* 2000). In this work, we assume that the mobile phone is carried by a user in his pant pocket. Most of usual daily activities involve movement of legs, therefore the best place to position the mobile phone is the user's pant pocket. This assumption is consistent with the large scale empirical study (Ichikawa *et al.* 2005) of 419 users conducted by Nokia across 3 different cities, which showed that majority of users either carry the mobile phone in their pant pocket (34%) or hand bag (33%). Carrying the phone in pant pocket also offers easy physical contact, good audibility and tactility, and close contact with the body which is useful when the phone is in silent-vibration mode. While our study assumes that the phone is carried in a pant pocket, a similar application can also be designed conveniently to suit the behaviours patterns of users who carry the smartphone in hand bag or backpack.

Different physical activities results in different movement patterns in data provided by acceleration sensors, and thus can be classified accordingly, using a machine learning algorithm. Figure 2 shows an example of acceleration signals (along x, y, and z axis) captured by acceleration sensors over time (samples). The data is captured over a consecutive sequence of walking and resting scenarios. Different activities (walking or resting) has been marked in the figure. As can be seen from the figure, there is a significant difference in pattern of acceleration for different activities. In this work, acceleration signal samples are sent to a server for further analysis. However it is also possible to carry out the analysis on the mobile device as well.

## *3.2 Preprocessing*

A change in acceleration pattern captured by the smartphone is caused due to the movement of the smartphone. While the acceleration pattern is mainly correlated with physical activity of the user, such patterns can also be corrupted due to "acceleration noise". Consider the case of a person being inside a car, train or lift. In such scenarios, apart from sensing the acceleration pattern of the user itself, the accelerometer also records acceleration changes due the movement of the car, train or lift.

As we are interested in analysing user physical activities, other acceleration sources such as those due to vehicular movement should be filtered out. According to our studies, acceleration pattern caused by physical activities has higher frequency content, while other sources such as vehicle and gravity force result in lower frequency contribution. For pre-processing step, we have used traditional signal processing approach of using a high pass filter to remove low frequency components and preserve high frequency components which are more representative of actual user activity. This also removes the constant component which is due to gravity force. The high pass filter is applied on x, y, and z acceleration signals.

## *3.3 Feature extraction*

Once we have extracted the high frequency component of the accelerometer, we are then interested to intelligently interpret this information. This information can be used to estimate the actual level (amount) of user physical activity, as well as classifying the activities into basic movements like walking, running, resting, and no activity (e.g. mobile phone is left on a table). In order to achieve this goal, we extract certain features from acceleration signals which represent different activities in a discriminative way.

The accelerometer continuously sample the acceleration experienced by the smartphone at each sampling interval and generates a 3-D acceleration vector along each of the x, y and z axis. Acceleration samples values changes along each of these axes are represented by $a_x$, $a_y$, and $a_z$ respectively. Changes in the value of $a_x$, $a_y$, and $a_z$ is dependent on the phone orientation, and will therefore vary with changes in physical orientation of the smartphone. While there will be limited movement in the physical orientation of the phone inside the user pant pocket, nevertheless to compensate for changes in the physical orientation of the phone we have used absolute magnitude of acceleration, as well as the rate of change in absolute magnitude acceleration as features. The mean of the absolute acceleration and rate of change in acceleration over a predefined window is also used for feature extraction. Essentially we aim to measure the acceleration as a scalar quantity rather then as a vector quantity. Absolute magnitude of acceleration at a sample is defined as

$$a = \sqrt{(a_x^2 + a_y^2 + a_z^2)}$$

Rate of change in acceleration (also known as jerk in physics) is defined as the difference between absolute magnitude acceleration for current sample and previous one.

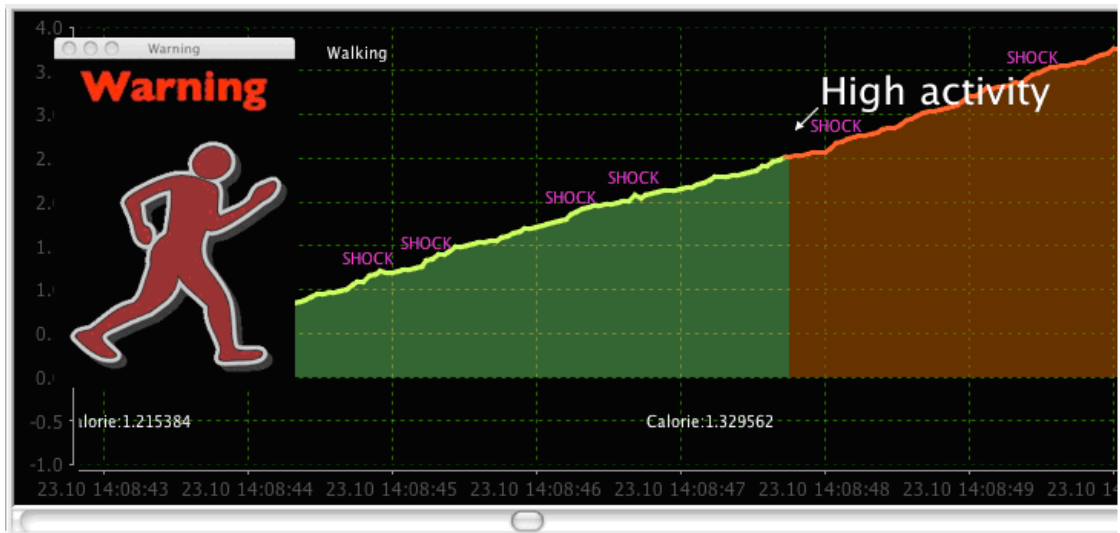

Figure 3. Warning in case of unexpected events

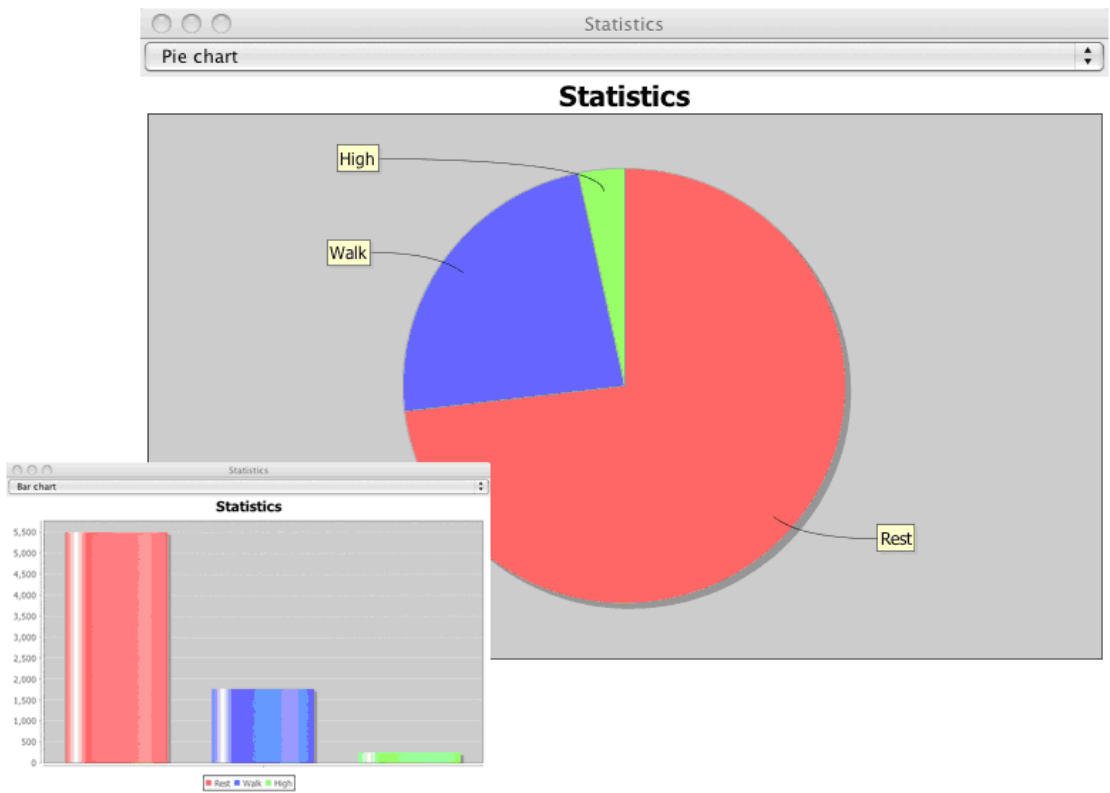

Figure 4. Visualizing different statistics on health related factors.

### *3.4 Estimating activity level*

One of the interesting factors for monitoring a user is his level (amount) of physical activity. The user can be provided warning in an event of high activity over a predefined period as shown in Figure 3. Acceleration magnitude (a) is correlated with activity level, however due to movements of legs, acceleration pattern comes with high frequency oscillations. In order to extract activity level, we measure the absolute difference between a pick in acceleration magnitude and subsequent valley. Activity level estimates can be presented to a monitoring person (agent) at a remote server side. A history of activity level estimates can be also stored for browsing and analysis by a medical staff member/ doctor.

Monitoring user activities can be further facilitated if the system on the server side is able to assist the monitoring person (agent) by classification of activities. In this way, a semi-automatic monitoring scheme can be applied. This means that monitoring by a person can be applied only in case of risky user activities, or if the activity level extends below or above a threshold (over certain period of time). In the next section, we describe our approach for automatic classification of user activities.

### *3.5 Activity classification*

As mentioned earlier, in addition to monitoring user physical activity level, we are also interested in the classification of user activities. This facilitates analysis and browsing of user activities by the monitoring agent at the server side, and also helps the agent to detect risky and emergency scenarios related to certain users. The agent can choose to browse activities belonging to a certain class, search for certain events or activities, or get alarm in case of certain activities. Additionally, the agent can monitor several users when he is assisted by an activity classification system. Since activity classification can send alerts, monitoring can be limited only to cases which a user is engaged in a risky activity. Therefore, the agent needs less concentration for monitoring and would be able to monitor several users simultaneously. Such activity classification application can also be useful in providing user statistical information such as the one given in Figure 4 for personal healthcare purposes for example.

In this work, we classify user activities in 4 main classes: walking, running, resting, and no activity. We build reference statistical models for these classes during a training phase. The statistical models are built using features extracted from acceleration signals.

As statistical model, we have used Gaussian mixture models - GMMs (McLachlan and Basford 1988). For each class, a GMM is trained to maximize likelihood of instances for that class:

$$\hat{\theta}_i = \arg\max_{\theta_i} p(X_i | \theta_i)$$

Where $\theta_i$ is the set of parameters of GMM for class, which is adjusted during training to maximize the likelihood and obtain $\theta_i$. Maximization of likelihood is done using Expectation-Maximization (EM) algorithm (Bilmes 1998).

During the test of the system, the trained GMM models are matched against actual instances of acceleration based features.

The activity class which maximizes likelihood is selected as ongoing activity class of user:

$$\hat{i} = \arg\max_i p(X | \theta_i)$$

Where $i$ is the selected activity class (result of classification). The activity classification results can be presented to the agent along with the estimation of activity level for a user by marking different activity diagrams with different colours.

The activity class label can be also stored along with activity level, in order to allow the agent to browse/search activity data later.

In the following section, we present our initial experiments for user authentication based on data captured by a mobile device (audio and motion) during regular physical activities. We show that users can be classified with high accuracy based on captured information using a mobile phone in their pant pocket.

## 4. Experimental studies and results

### 4.1 Physical activities classification

We set up initial experiments for estimating user physical activity level classifying these activities. We have used Apple iPhone 3G as the mobile phone for the experiments. Linear acceleration signals are provided along x, y, and z axis by the iPhone accelerometer at 5 Hz rate. We recorded a database of 320 activity instances with 4 subject users. The database is portioned into 208 activity instances for training and 112 activity instances for testing the system. There are four activity classes: walking, running, resting, and no activity (e.g. mobile phone on a table). During resting, the user is either using a laptop or watching TV while sitting. Each activity class has equal number of instances in the database. Each activity instance lasts 10 seconds. The subject users carry the mobile phone in their pocket in a regular manner during different activities.

The acceleration signal is preprocessed as explained in Section 3.2. Features vectors are extracted for every sample of acceleration signal. The activity level estimation step is done by measuring absolute difference between consecutive picks and valleys in acceleration magnitude.

For activity classification, values of features are averaged over activity instance interval (10 seconds in this case), resulting in an average feature vector for every activity instance. Extracted features are used to train GMMs for each class. As mentioned before, we are interested in classifying user activities into 4 activity classes - walking, running, resting and no activity. We have used 2 Gaussian mixtures for each class. The parameters of Gaussians are trained using expectation-maximization (EM) algorithm to maximize likelihood for each class. While testing the system, extracted features are matched against models for different classes. Each class model provides a likelihood score indicating the match between the actual activity instance and the model. Therefore we obtain 4 likelihood scores for each activity instance. The class having the highest likelihood score is selected as the outcome of activity classification.

Table 1. Confusion matrix for different activities.

| Activity | Walk | Run | Rest | No Act. |
|---|---|---|---|---|
| Walk | 26 | 2 | 0 | 0 |
| Run | 2 | 26 | 0 | 0 |
| Rest | 0 | 0 | 25 | 3 |
| No Act. | 0 | 0 | 1 | 27 |

We have evaluated the activity classification system in terms of the accuracy in detection of activities. The overall accuracy is **92.9%**. Table 1 shows a confusion matrix for the errors. This table indicates which classes are mostly confusable. For instance, we can see that a walk activity instance is detected 26 times as walking, 2 times as running, and never as resting or no activity class. According to the table, confusion between walking and running classes, also between resting and no activity classes is higher.

*4.2 Unexpected event detection*

For unexpected event detection, the experimental setup is similar to (Ketabdar 2009). We have used iPhone 3G for the experiments. Acceleration and audio data is recorded using embedded microphone and sensors through a data collection application developed for iPhone. For the experiments, we have recorded a database of normal and risky (as defined) situations. In this database, there are 98 samples of normal situations, and 36 samples of physical shocks. In order to obtain physical shock, we let the iPhone to fall on a carpet or wooden floor form a distance of approximately 75 cm. In order to obtain normal condition samples, we let 5 users to carry the iPhone normally in their pocket, hand or bag for a period of 10 seconds. These users indulge in different day-to-day activities such as walking, jogging, taking lift and walking on stairs. We tried to have different variety of scenarios, especially those which can have similarity to a shock (due to high physical activity) such as walking on stairs and taking lift. In this way, we can make sure that our algorithm is able to distinguish between such cases and a real risky shock (Ketabdar 2009).

As mentioned earlier, the risky situation is defined as a sequence of free fall, impact (shock), and no activity period. The free fall is detected when the norm of acceleration signals (along x, y, and z directions) falls below a predefined threshold. The no-activity period is identified when the average of norm of acceleration signals in an interval after the impact (8 seconds in our case) falls below a threshold. The impact (shock) situation is detected by comparing features extracted from acceleration and audio data against a statistical model created for impact (shock). The model which is used for this work is a Multi-Layer Perceptron (MLP) trained using samples of shock (impact) and regular situation collected as previously mentioned. The features used in this study are mainly based on average and variance of acceleration components (as well as their norm), and audio signal. The MLP is then able to classify new samples of features as shock or normal (regular) situation. The "risky situation" is detected upon detection of free fall, shock, and period of no-activity in correct order. Table 2 summarizes initial results. Our studies show that defining the three steps for risky events detection can significantly reduce number of false alarms (Table 2). The first row in the table shows the results when the three step definition is used, and the second row shows results when only impact is considered as risky event.

Table 2. Results for detection of a risky situation which can lead to having a mobile device being lost or stolen.

| Algorithm | Accuracy | True Alarm | False Alarm |
|---|---|---|---|
| *3 step definition* | **94.4** | 34 | 4 |
| *Only impact* | **86.1** | 31 | 9 |

*4.3 Implicit user authentication*

We set initial experiments to investigate possibility of implicit user authentication based on regular physical activities of user (walking in our case).

For experiments, we recorded device motion information (using embedded acceleration sensors) as well as ambient audio (using embedded microphone). The recoding is done during regular physical activities which are walking in this case. The device is carried in user's pant pocket. We have used iPhone as mobile device and we recorded the signals using a data collection application we developed for iPhone. We have invited 9 participants for the experiments. We captured audio at 8 KHz and acceleration at 50 Hz using embedded sensors in the iPhone. We let the iPhone to be placed regularly in the pocket, without fixing its position or orientation. The test users are asked to walk for about 2 minutes in indoor and outdoor environments. The recording for each user is repeated over 3 different days. Users were asked to come for the experiments with different sets of shoes and pants in different days, in order to take into account the effect of variability in clothing in the identification process. Feature extraction is the first processing step. We extract two sets of features, one from acceleration signals and one from audio signal. Features are extracted over a window of 2 seconds of acceleration and audio signals. For acceleration signals, the extracted features are mainly based on average, variance and magnitude of acceleration components.

Here is a list of features:
• Average field strength along x, y, and z directions.
• Variance field strength along x, y, and z directions.
• Average of Euclidian norm of filed strength along x, y, z.
• Variance of Euclidian norm of field strength along x, y, and z.
• Piecewise correlation between field strength along x and y, x and z, and y and z.

Table 3. User identification results using different feature sets (movement, audio, movement+audio)

| Feature source | Accuracy |
|---|---|
| *Movement* | 88.3 |
| *Audio* | 47.8 |
| *Movement + Audio* | **90.1** |

Table 4. User authentication measures for some users.

| User ID | Precision | Recall | F-measure | ROC Area |
|---|---|---|---|---|
| *1* | 0.89 | 0.95 | 0.92 | 0.98 |
| *2* | 0.92 | 0.87 | 0.90 | 0.96 |
| *3* | 0.92 | 0.91 | 0.92 | 0.98 |
| *4* | 0.92 | 0.92 | 0.92 | 0.97 |
| *Weighted Average* | **0.91** | **0.91** | **0.91** | **0.97** |

For audio signal, extracted features are mainly based on average, variance, and energy of the audio signal in each window. Variance of Fourier transform of audio signal is also used as a feature.

Extracted features are feed as input to MLP for user classification/ identification. Table 3 shows classification results for different feature sets. We report results for using movement (acceleration) based features, audio based features, and a combination of movement and audio based features. As can be seen from the table, the combination of audio and movement based features provide the best user identification results (90.1%). Table 4 shows identity verification (authentication) measures for some of the users. The Receiver Operating Characteristic (ROC) measurements show a good tradeoff between true and false alarms indicating significant user authentication results.

In this experiment, we have presented initial results for user authentication over a window period of 2 seconds. This means that every 2 seconds, we are able to re-authenticate the user. However, such a short interval continuous re-authentication may not be necessary in practical applications. It may be enough to have an authentication measure for instance, every minute. In such a case, short interval (2 seconds window) based authentication results can be used in a voting scheme. The identified user over a minute is the user having highest vote (recognition) in 2 second based windows. Our experiments show that user identification accuracy in this case rises to 97.5% using combination of audio and acceleration based features.

## 5. *ActivityMonitor:* The Developed System

We have developed a system called as "ActivityMonitor" based on the idea presented in this paper for live remote monitoring of several users physical activities. In this section, we explain the setup and functionalities of this system.

In order to setup and operate the system, two applications are required. The first application is installed and executed on an iPhone. This application sends acceleration data through available data service (Wi-Fi, GPRS, etc.) to a designated server. The user can set certain configuration parameters such as a name for ongoing experiment and sampling frequency. In addition, the application allows capturing data in snapshots in order to reduce traffic of the server.

The second application is a Java application (ActivityMonitor) which can be installed on any ordinary computer. This application connects to the designated server and reads acceleration data. The data is then analyzed and presented (to an agent) in real time as physical activity information, and different statistics and health related factors (Figure 5). The application is able to classify activities in certain categories, and issue warnings in case of irregular activity patterns. It can additionally store the data, browse it, or search for certain activity category. The ActivityMonitor screen has different sections. The main part of the screen is allocated by plots of activity related data. In these plots, activity level and category of the user, warnings, and energy (calorie) consumption can be visualized (Figure 6.a). The ActivityMonitor screen also comes with a log field on the left side, which provides textual information about the ongoing activity and different statistics (Figure 6.b). There is also a settings tab (Figure 3.c) which allows configuring server connection, managing data download, and formatting the data. The monitoring agent can also choose to observe statistical data in a separate window (Figure 3).

The desktop application can be used to monitor several users simultaneously. It can automatically analyze activity data and detect unexpected patterns such as shocks or long periods of high or low activity (Figure 4). Upon detection of an unexpected event for a certain user, the agent is informed by a visual or audio alert, and the monitoring screen related to that user pops up. This allows monitoring multiple users in an automatic or semi-automatic manner.

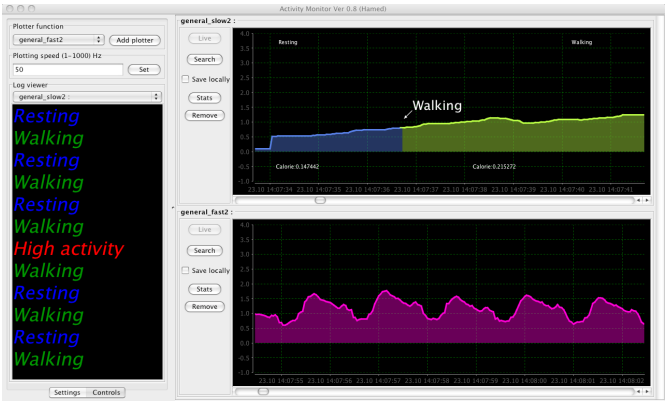

Figure 5. ActivityMonitor screenshot.

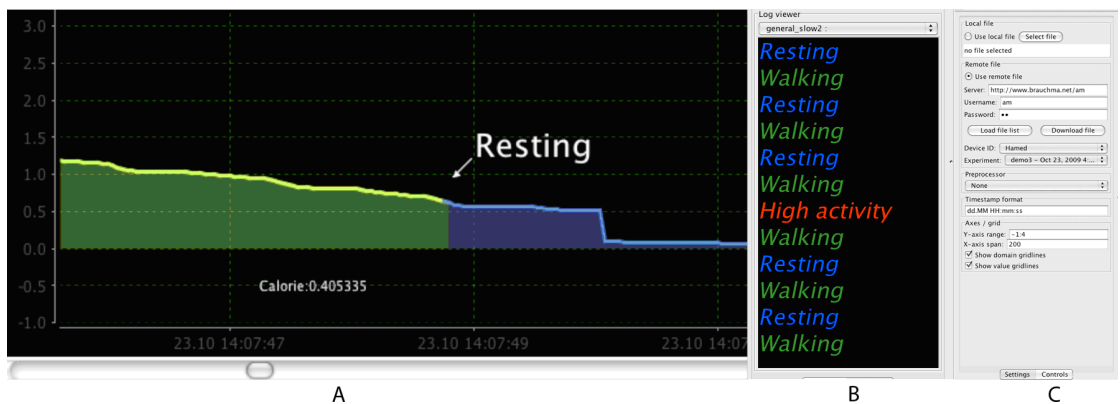

A B C

Figure 6. a) Main part, which visualizes activity of user, b) Log viewer which can show the on going activity class, warning etc. as text, c) Controls and settings part which can be used for adjusting different parameters in relation with the connection to server visualization etc.

## *5.1 Social Implications of ActivityMonitor*

As mentioned in the previous sections, ActivityMonitor system turns regular mobile devices to devices for constantly monitoring physical activities of users. Such a system can be provided to public as software for different mobile devices, as well as a server and monitoring centre. Users who may wish to be monitored, can register themselves in this service. The registration can be something similar to registration for advanced services such as MMS, VOIP, etc. By registering in this service and activating the respective software on the mobile phone, the user allows his activity information to be transferred to the server and be monitored by an agent. As mentioned before, the monitoring process can be done in a semi-automatic manner due to the fact that the desktop monitoring application can automatically check for unexpected patterns. Therefore, monitoring by a human agent can be necessary only if something unexpected happens. This allows possibility of monitoring several users by an agent simultaneously.

     As an alternative, the system can be provided in a private manner. This means that the desktop monitoring application can be also sold as software for personal use. In this way, two people (monitoring agent and the user), or a group of people can establish their own monitoring process based on a local server. For instance, one can personally take care of his/her elderly parents using such a system. In this case, a software can be downloaded and installed on mobile phone, and a second software

(desktop monitoring application) plus some space on a server should b purchased. All these steps can be done online very efficiently which means a private monitoring system can be established within a few minutes.

Although such a system can be useful for healthcare purposes, as studied in Kargl *et al.* (2005), it can come with some privacy issues. Simply, people may not feel comfortable with having their activities always being monitored. Although this can be an important issue to be studied deeply before commercializing such systems, there are already some potential solutions. For instance, the user may simply switch of the monitoring application on his mobile phone, or leave it unattended when he does not want to be monitored. Another solution could be designing the monitoring software in way that only very general information about ongoing activity of the user such as activity level can be transferred. This allows monitoring the user and helping him in case of unexpected events, and in the same time those not expose detailed information to the agent.

## 6. Conclusion

In this paper, we have presented a system and methodology based on processing data provided by mobile phone accelerometer sensor for monitoring physical activities of users. As a mobile phone can be conveniently and constantly carried by a user, and it does not impose burden of wearing extra sensors, such an application can enable the mobile phone to become as a user friendly, precise and constant health and activity monitoring device.

We have shown novel accelerometer based applications which run on smartphones. These applications can be used for activity monitoring, and while it has been shown to benefit a wide range of users, it can in particular benefit those requiring healthcare administration and the elderly. Such application can be implemented without any additional hardware overhead, as accelerometers are embedded by default in the latest generations of smartphones, and allow third vendor application development and distribution through the smartphone vendor online application store.

In addition to monitoring activity level and activity classification which is already discussed in this paper, such a system can be used for more detailed analysis of certain activities. For instance, walking pattern of a user can be analysed to see if there is any deviation form the normal pattern of walking for the same user or users of the same age. Many diseases show their early symptoms in changes of daily activity pattern. Our system can be used for advanced analysis of certain activities and early detection of certain problems, as well as monitoring progress of user after a surgery or medical treatment.

We have further shown how the concept of analysing sensory data captured by a mobile device can be further extended to provide mobile phone security and authentication. This has been supplemented with our investigations on the correlation between motion and audio information captured using embedded sensors in a mobile device for enhancing security functionalities related to the device. We have mainly investigated two cases. In the first case, we detect unexpected events, based on audio and movement analysis. We showed that an unexpected event such as a phone fallen down and left unattended can be identified with a high accuracy. For the second case, we proposed implicit user identification/authentication based on audio and movement analysis captured by the user phone during regular physical activities (e.g., walking). We showed that a user can be identified with high accuracy on this basis. The results of such analysis can be used to arrange a graded security scheme for mobile and handheld devices based on their actual status. The proposed framework can be used as a stand-alone implicit security enhancement technique, or used as a complement to regular user authentication techniques. These results can be an initiation for a new security paradigm for enhancing security functionalities in mobile devices based on audio and movement analysis.